\documentclass[11pt,a4paper]{amsart}
\usepackage{amsthm,amscd,times,amssymb,amsmath,mathrsfs,hyperref}

\DeclareMathOperator{\Hom}{Hom}
\newcommand{\C}{\mathbb{C}}
\newcommand{\Aut}{\mathrm{Aut}}
\newcommand{\End}{\mathrm{End}}

\def\eps{\epsilon}

\def\be{\begin{equation*}}
\def\ee{\end{equation*}}
\def\bel{\begin{equation}}
\def\eel{\end{equation}}
\newcommand{\bea}{\begin{eqnarray}}
\newcommand{\eea}{\end{eqnarray}}
\newcommand{\beas}{\begin{eqnarray*}}
\newcommand{\eeas}{\end{eqnarray*}}

\begin{document}

\include{trees}

\title[Nonperturbative Renormalization of Schwinger--Dyson D-Module]{Nonperturbative Renormalization as Riemann--Hilbert Decomposition\ of Schwinger--Dyson D-Module}
\author{Chaoming Song}
\email{c.song@miami.edu}
\address{%
Department of Physics, University of Miami, Coral Gables, Florida 33146, USA}%

\begin{abstract}
We present a non-perturbative formulation of renormalization by viewing the regularized Schwinger–Dyson hierarchy as a meromorphic connection, that is, as a \( \mathscr D \)-module on the product of space–time with the regulator disc.  The irregular Riemann–Hilbert correspondence splits this connection into a purely formal submodule that contains every ultraviolet pole and a holomorphic submodule that is finite at the regulator origin. In this setting, counterterms coincide with the formal Stokes data, the renormalized theory is identified with the analytic submodule, and the Callan–Symanzik flow appears as an isomonodromic deformation in the physical scale. Viewed through this lens, the Connes–Kreimer construction with its graph Hopf algebra, Bogoliubov recursion, and Birkhoff factorisation is simply the perturbative shadow cast by the global geometric decomposition of the full Schwinger–Dyson system.
\end{abstract}

\maketitle

\section{Introduction}

Renormalization was first formalised perturbatively, where ultraviolet divergences are removed order by order in the coupling by adding local
counterterms.  The most systematic realisation of this idea is due to
Connes and Kreimer, who interpreted perturbative renormalization as a
Birkhoff factorisation in the loop group
\(\Hom(H,\mathbb{C}((\epsilon)))\) of characters on the Hopf algebra
\(H\) of Feynman graphs \cite{connes1999hopf,connes1999renormalization,connes2000renormalization,connes2001renormalization}.
Their framework unifies the Bogoliubov recursion, the forest formula,
and the scale dependence of Green functions, but it remains tied to
power series in the coupling and therefore does not address the
non-perturbative content of quantum field theory.

A complementary programme treats quantum fields through their Schwinger–Dyson (SD) equations, an infinite hierarchy of functional identities that fix all Green functions once a set of renormalization
conditions is supplied \cite{dyson1949s,schwinger1951greenI}.  These equations are exact and in principle non-perturbative, yet in practice they are difficult to close.  Truncations often break gauge invariance, while diagrammatic expansions re-introduce the same infrared and ultraviolet complications that the SD hierarchy was
meant to avoid. 

Several authors have tried to combine the algebraic control of
Connes–Kreimer with the functional content of the SD hierarchy.
Bergbauer and Kreimer showed that a single Dyson--Schwinger equation can be rewritten as a fixed-point equation inside the graph Hopf algebra, thus linking the combinatorial and functional approaches but still remaining within the perturbative domain~\cite{bergbauer2006hopf,manchon2008hopf}.  More recent works have pushed these ideas further~\cite{balduf2024dyson}, seeking to extract non-perturbative information~\cite{kreimer2006etude}, yet a fully non-perturbative treatment of the Schwinger--Dyson hierarchy remains elusive.

In this paper we take a different route.  We regard the regularized SD
hierarchy itself as a meromorphic connection, that is, as a
\(\mathscr{D}\)-module on the product of space–time with the regulator
disc.  The irregular Riemann–Hilbert correspondence then splits this
connection into a formal submodule that carries all ultraviolet poles
and a holomorphic submodule that is finite at the regulator origin.
Counterterms become formal Stokes data; the renormalized theory is
identified with the analytic submodule; the Callan–Symanzik flow is an
isomonodromic deformation in the physical scale.  When the resulting
factorisation is expanded in the coupling, one recovers exactly the
Connes–Kreimer Hopf algebra, the Bogoliubov recursion, and the Birkhoff
factorisation.  The Connes–Kreimer construction therefore appears as
the perturbative shadow of a geometric decomposition that exists
before any expansion and applies to the entire SD hierarchy at once.

\section{Schwinger–Dyson D-Module}

The Schwinger–Dyson equations form a hierarchy of ODEs given by:
\[
  \mathcal{D}_x
  = \frac{\delta S[J]}{\delta \phi(x)} \Bigg|_{\phi \mapsto \frac{\delta}{\delta J}},
\]
where \( S[J] = \int d^{D} x \, \left( \mathcal{L}(\phi, g_0) + J(x)\phi(x) \right) \) is the action, \( \mathcal{L}(\phi, g_0) \) is the Lagrangian density, \( J(x) \) is an external source, and \( g_0 \) is the bare coupling constant. The generating functional \( Z[J] \) satisfies the equation:
\[
  \mathcal{D}_x \, Z[J] = 0 \quad \forall x.
\]
The D-module \( \mathcal{M} \) is the solution space to the Schwinger–Dyson equations:
\[
\mathcal{M} = \frac{\mathcal{D}}{\langle \mathcal{D}_x \rangle}.
\]
Here and throughout, we use ``D-module'' in the sense of a coherent module over a ring of differential or pseudo\-differential operators, as discussed in~\cite{kashiwara2003d,hotta2007d}.

This definition is formal, but the system suffers from singularities, which require regularization. We introduce \( \eps \) as a regularization parameter and \( \mu \) as the physical scale, where \( \eps \) controls the UV cutoff. The bare coupling constant \( g_0(\eps) \) is not a fixed parameter but instead is treated as a Laurent series in \( \eps \). The simplest regularization method is point-splitting regularization, where \( \eps = -\ln r_{\text{UV}} \).  

For more sophisticated dimensional regularization, we use \( d = D - \eps \), which introduces a smooth regularization that mitigates the singularities. This prevents the direct application of differential operators; instead, we must consider pseudo-differential operators and employ microlocal analysis. Despite the technical complications, the D-module can be defined in the microlocal sense~\cite{kashiwara2003d}, as we discuss in Appendix~\ref{sec:dim_reg}.

Accordingly, the Schwinger–Dyson operators, which govern the Green's functions, are defined as:
\[
  \mathcal{D}_x(\eps, \mu) 
  = \frac{\delta S_{\eps, \mu}[J]}{\delta \phi(x)} \Bigg|_{\phi \mapsto \frac{\delta}{\delta J}}.
\]
The generating functional \( Z[J] \) satisfies:
\[
  \mathcal{D}_x(\eps, \mu) \, Z[J] = 0 \quad \forall x.
\]
The D-module \( \mathcal{M} \) is the solution space to the Schwinger–Dyson equations:
\[
  \mathcal{M} = \frac{\mathcal{D}_{\eps,\mu}}{\langle \mathcal{D}_x(\eps, \mu) \rangle},
\]
which defines a meromorphic connection in the \( \eps \)-direction, where \( \eps \) represents the regularization parameter. This connection has an irregular singularity at \( \eps = 0 \).

Promote the scale to a coordinate on the punctured multiplicative line
\( \C_\mu^{\!*}=\{\mu\neq0\} \)
and extend the base further to
\( X\times \partial_\eps\times\C_\mu^{\!*} \).
The total connection decomposes as
\[
  \nabla \;=\;
    \sum_{\nu=1}^{D}
       \nabla_{\partial_{x^\nu}}\,\mathrm{d}x^\nu
    \;+\;
       \nabla_{\partial_\eps}\,\mathrm{d}\eps
    \;+\;
       \nabla_{\mu\partial_\mu}\,\frac{\mathrm{d}\mu}{\mu},
\]
with
\[
  {\;
      \nabla_{\partial_\eps}
      \;=\;
      \partial_\eps
      + A(\eps,\mu)},
  \qquad
  {\;
      \nabla_{\mu\partial_\mu}
      \;=\;
      \mu\partial_\mu
      + B(\eps,\mu)} .
\]
Here
\(A(\eps,\mu)=\displaystyle\sum_{k\ge1}\frac{A_k(\mu)}{\eps^{\,k}}\)
collects the UV–pole structure,
while \(B(\eps,\mu)=\gamma(\mu)+O(\eps)\)
is regular at \( \eps=0 \) and encodes the
anomalous–dimension / beta–function matrix.  
Flatness of the full connection imposes
\(
   [\nabla_{\partial_\eps},\nabla_{\mu\partial_\mu}]=0
\),
which is precisely the Callan–Symanzik compatibility
between RG flow and the subtraction of \(1/\eps\) divergences.

\section{Irregular Riemann–Hilbert Decomposition}

Consider the punctured regulator disc \(0<|\eps|<R\).  
For the Schwinger–Dyson D-module \(\mathcal{M}\) every fundamental solution  
\[
   \gamma(\eps,\mu):\mathcal{M}\longrightarrow\C((\eps)),
   \qquad
   \nabla\,\gamma(\eps,\mu)=0 ,
\]
provides a basis of flat sections along the \(\eps\)-direction.  
Explicitly,
\[
   \nabla_{\partial_\eps}
      \;=\;
      \partial_\eps + A(\eps,\mu),
   \qquad
   A(\eps,\mu)=\sum_{k\ge1}\frac{A_k(\mu)}{\eps^{\,k}}
\]
is meromorphic with a finite-order pole at \(\eps=0\).

\medskip
\noindent\textbf{Theorem (Levelt–Turrittin~\cite{turrittin1955convergent,levelt1961hypergeometric}).}  
After a finite ramified cover \(\eps=t^{m}\), any meromorphic connection admits a \emph{formal} decomposition
\[
   (\mathcal{M}\!\otimes\!\C((t)),\nabla)
   \cong
   \bigoplus_i
      \exp\!\bigl(Q_i(t^{-1})\bigr)\,
      t^{R_i}\!\otimes\!\widehat{\mathcal{R}}_i,
\]
where \(Q_i\) are polynomials, \(R_i\) residue matrices, and
\(\widehat{\mathcal{R}}_i\) regular–singular factors.
Equivalently, a unique formal gauge
\[
   \gamma_-(\eps)\in
   \Aut\!\bigl(\mathcal{M}\otimes\C((\eps))\bigr)
\]
brings \(\nabla\) to formal normal form, defining the \emph{formal submodule}
\[
   \mathcal{M}_-
       \subset
   \mathcal{M}\otimes_{\C\{\eps\}}\C[[\eps^{-1}]] .
\]

\medskip
\noindent\textbf{Malgrange–Sibuya condition.}  
We say that \(\nabla\) satisfies the Malgrange–Sibuya (MS) condition if, in every sufficiently small sector at \(\eps=0\), there exists an \emph{analytic gauge}
\[
   \gamma_+(\eps)\in
   \Aut\!\bigl(\mathcal{M}\otimes\C\{\eps\}\bigr),
   \qquad
   \nabla^+=\gamma_+\,\nabla\,\gamma_+^{-1},
\]
such that \(\nabla^+\) is holomorphic at \(\eps=0\).
Sectorial mismatches of \(\gamma_+\) encode the usual Stokes data and yield the \emph{analytic submodule}
\[
   \mathcal{M}_+
       \subset
   \mathcal{M}\otimes_{\C\{\eps\}}\C\{\eps\}.
\]

\medskip
\noindent\textbf{renormalizability.}  
A quantum field theory is called \emph{renormalizable} if, after adding a \emph{finite} set of additional relevant operators to the Lagrangian, its Schwinger–Dyson D-module satisfies the MS condition.  
If no such finite enlargement works, the theory is \emph{non-renormalizable}.

\medskip
\noindent\textbf{Theorem (Malgrange–Sibuya~\cite{malgrange1991equations,sibuya2008linear}).}  
Every finite-rank meromorphic connection fulfils the MS condition automatically.  

Consequently, ordinary perturbatively renormalizable local QFTs—whose divergent operator basis is finite—belong to this class (Appendix \ref{sec:finite_rank_perturbative}).  
For infinite-rank D-modules the MS condition singles out exactly those theories whose UV behaviour can still be controlled; when it fails, couplings evolve non-analytically and additional divergences appear.

\medskip
\noindent\textbf{Gluing of submodules.}  
When the MS condition holds (possibly after enlarging the coupling space) the formal and analytic parts glue over the punctured disc:
\[
   \mathcal{M} \;=\; \mathcal{M}_- \cap \mathcal{M}_+,
   \qquad
   \gamma(\eps,\mu)
        = \gamma_-^{-1}(\eps,\mu)\,\gamma_+(\eps,\mu).
\]

\medskip
\noindent\textbf{renormalized D-module.}  
The fibre at \(\eps=0\),
\[
   {
      \mathcal{M}_{\mathrm{ren}}
         = \mathcal{M}_+\big|_{\eps=0}
         = \mathcal{M}_+ / \eps\,\mathcal{M}_+
   },
\]
inherits a pole-free connection.  It therefore supplies: i) regular ordinary differential equations in \(\eps\) (no \(1/\eps\) terms), ii) finite Callan–Symanzik equations in the scale \(\mu\), iii) a closed divergence-free hierarchy of renormalized Schwinger–Dyson relations.

\section{Connections to Connes–Kreimer’s Approach}

To make contact with Connes–Kreimer’s Hopf–algebraic formulation of perturbative
renormalization, fix the \emph{perturbative fundamental solution}
obtained by expanding every Green function as a formal power series in
the coupling.  In this basis the objects introduced above match the
ingredients of Connes–Kreimer theory as follows:

\noindent\textbf{Divergent graphs vs.\ formal submodule.}  
The Hopf algebra \(H\) of Feynman graphs is generated by one–particle–irreducible diagrams; its coproduct \(\Delta\Gamma=\sum_{\gamma\subset\Gamma}\gamma\otimes\Gamma/\gamma\) organises all nested subdivergences. The vector space \(H\otimes\C[[\epsilon^{-1}]]\) of Laurent series in \(\epsilon^{-1}\) corresponds exactly to the \emph{formal submodule} \(\mathcal{M}_-\subset\mathcal{M}\otimes_{\C\{\epsilon\}}\C[[\epsilon^{-1}]]\), because every coefficient of a pole term can be expanded in that graph basis.

\noindent\textbf{Counter-term character vs.\ formal gauge \(\gamma_-\).}  
A character \(\phi_-:H\to\C[[\epsilon^{-1}]]\) implementing minimal subtraction assigns to each graph the \emph{counter-term} that cancels its pole part. Acting on \(\mathcal{M}_-\) by left multiplication, \(\phi_-\) is identified with the formal gauge \(\gamma_-\in\Aut\bigl(\mathcal{M}\otimes\C((\epsilon))\bigr)\) provided by the Levelt–Turrittin theorem. Its sole effect is to annihilate all negative powers of \(\epsilon\).

\noindent\textbf{renormalized character vs.\ analytic submodule \(\mathcal{M}_+\).}  
The Rota–Baxter projection \(R_+\colon\C((\epsilon))\to\C\{\epsilon\}\) extracts the pole-free part of every Laurent series. Acting diagram-wise on the bare amplitude \(\phi\in\Hom(H,\C((\epsilon)))\), one obtains the \emph{renormalized character} \(\phi_+:=R_+\circ\phi\), which is holomorphic at \(\epsilon=0\). The image of \(\phi_+\) is precisely the \emph{analytic submodule} \(\mathcal{M}_+\subset\mathcal{M}\otimes_{\C\{\epsilon\}}\C\{\epsilon\}\).

\noindent\textbf{Birkhoff decomposition vs.\ Riemann–Hilbert decomposition.}  
Connes and Kreimer proved that every bare character factorises uniquely as \(\phi=\phi_-^{-1}\star\phi_+\) in the \emph{loop group} \(\Hom(H,\C((\epsilon)))\), a statement known as the \emph{Birkhoff (or Bogoliubov–Parasiuk) decomposition}. This exactly mirrors the intrinsic \emph{irregular Riemann–Hilbert decomposition} obtained above, where \(\gamma(\epsilon,\mu)=\gamma_-^{-1}(\epsilon,\mu)\gamma_+(\epsilon,\mu)\), and the convolution product matches the composition of gauges.

\medskip

In a nutshell, Connes–Kreimer’s Hopf–algebraic renormalization appears as the
\emph{loop-group avatar} of the Riemann–Hilbert decomposition of the
Schwinger–Dyson D-module: the coproduct encodes the SD Leibniz rule,
\(\gamma_-\) is the counter‐term character, \(\gamma_+\) the renormalized
character, and the Birkhoff decomposition mirrors the gluing
\(
  \mathcal{M}=\mathcal{M}_-\cap\mathcal{M}_+
\)
performed in the present framework.

\section{Extracting the Renormalized ODEs}

Before subtraction the $\epsilon$–component of the Schwinger–Dyson
connection acts on the full module $\mathcal{M}$ as
\[
   \nabla_{\partial_\epsilon}
   \;=\;
   \partial_\epsilon+A(\epsilon,\mu),
   \qquad
   A(\epsilon,\mu)
     =\sum_{k=-N}^{\infty}A_k(\mu)\,\epsilon^{\,k}
     \;\in\;
     \End(\mathcal{M})\otimes\C((\epsilon)).
\]
The coefficients with $k<0$ are the ultraviolet poles.

\medskip
\noindent\textbf{Pole–killing gauge.}
By the Levelt–Turrittin theorem there exists a \emph{formal gauge}
\[
   \gamma_-(\epsilon,\mu)
   \;\in\;
   \Aut\!\bigl(\mathcal{M}\otimes\C[[\epsilon^{-1}]]\bigr)
\]
that eliminates every negative power of~$\epsilon$.  Conjugating the
connection gives the \emph{holomorphic} form
\[
   \nabla^{+}_{\partial_\epsilon}
   =\gamma_-\,\nabla_{\partial_\epsilon}\,\gamma_-^{-1}
   =\partial_\epsilon+A^{+}(\epsilon,\mu),
   \qquad
   A^{+}(\epsilon,\mu)
     =\sum_{k\ge0}A^{+}_{k}(\mu)\,\epsilon^{\,k}
     \in\End(\mathcal{M}_{+})\otimes\C\{\epsilon\}.
\]

\medskip
\noindent\textbf{Regular $\epsilon$–ODE.}
For any flat section $\Psi(\epsilon,\mu)$ of the renormalized
connection we obtain the \emph{regular} ordinary differential equation
\[
   \frac{\partial}{\partial\epsilon}\Psi(\epsilon,\mu)
   \;+\;A^{+}(\epsilon,\mu)\,\Psi(\epsilon,\mu)=0,
\]
which at $\epsilon=0$ specialises to
\[
   \left.\partial_\epsilon\Psi\right|_{\epsilon=0}
   =-\,A^{+}_{0}(\mu)\,\Psi(0,\mu).
\]

\medskip
\noindent\textbf{Scale component.}
If the original connection contains a scale piece
\(
   \nabla_{\mu\partial_\mu}=\mu\partial_\mu+B(\epsilon,\mu),
\)
the same gauge produces
\[
   \nabla^{+}_{\mu\partial_\mu}
   =\mu\partial_\mu+B^{+}(\mu),
   \qquad
   B^{+}(\mu)
     =\Bigl[\,
        \gamma_-B\gamma_-^{-1}
        +(\mu\partial_\mu\gamma_-)\gamma_-^{-1}
      \Bigr]_{\epsilon=0}.
\]
Flatness in the scale direction therefore becomes the
renormalization–group ODE
\[
   {\;
      \mu\,\frac{\mathrm{\partial}}{\mathrm{\partial}\mu}\,
      \Psi(0,\mu)
      =-\,B^{+}(\mu)\,\Psi(0,\mu)
   \;}
\]
inside the renormalized module $\mathcal{M}_{\mathrm{ren}}$.

\medskip
\noindent\textbf{Perturbative (Bogoliubov) route.}
Choosing a formal power–series ansatz in the coupling
\(g\) and expanding the gauge
\(
  \gamma_-=\mathbf 1+\sum_{n\ge1}g^{\,n}\gamma_-^{(n)}
\)
recovers the usual \emph{Bogoliubov recursion} on the Hopf algebra of
Feynman graphs: each $\gamma_-^{(n)}$ is the counter-term character
determined by minimal subtraction, while
\(
  \gamma_+=\gamma_-^{-1}\big|_{\text{finite}}
\)
produces the renormalized amplitudes.  In this
perturbative basis the entries of $A^{+}_{0}$ and $B^{+}$ are the
anomalous dimensions and $\beta$–function, so that the
$\mu$–ODE above becomes the familiar Callan–Symanzik equation
\[
   \bigl(\mu\partial_\mu
         +\beta(g)\,\partial_g
         +n\,\gamma(g)
   \bigr)\,
   G^{(n)}(\{p_i\};\mu)=0,
\]
now recognised as the scale component of the pole-free connection in
the renormalized D-module.

\section{Conclusion and Outlook}

We have reformulated renormalization as an irregular Riemann–Hilbert
decomposition of the Schwinger–Dyson D-module.  A single meromorphic
connection
\(
   \nabla_{\partial_\epsilon}=\partial_\epsilon+A(\epsilon,\mu)
\)
is split by a formal gauge
\(
   \gamma_-(\epsilon,\mu)\in\Aut\bigl(\mathcal{M}\otimes\C[[\epsilon^{-1}]]\bigr)
\)
that removes every pole and an analytic gauge
\(
   \gamma_+(\epsilon,\mu)\in\Aut\bigl(\mathcal{M}\otimes\C\{\epsilon\}\bigr)
\)
that is holomorphic at the origin.  Their gluing
\(
   \gamma=\gamma_-^{-1}\gamma_+
\)
yields a factorisation
\(
   \mathcal{M}=\mathcal{M}_-\cap\mathcal{M}_+
\)
in which counterterms appear as Stokes data, the fibre
\(
   \mathcal{M}_{\text{ren}}=\mathcal{M}_+/\epsilon\mathcal{M}_+
\)
is free of poles, and the scale component of the connection realises
the renormalization group as an isomonodromic deformation.  Connes and
Kreimer's Hopf-algebra factorisation is recovered by expanding these
objects in the coupling.

The Malgrange–Sibuya (MS) condition extracted from this geometry gives
a non-perturbative definition of renormalizability: a theory is
renormalizable precisely when the analytic gauge exists after adding
only finitely many relevant operators.  For finite-rank connections
this reproduces the textbook Bogoliubov counterterm construction; for
infinite rank it singles out the quantum field theories that remain
consistent beyond perturbation theory.

Two complementary approximation strategies emerge.  One may truncate
the module to a finite dimension and study the convergence of the
resulting matrices.  Alternatively, one may keep the full infinite
module and introduce a trans-series ansatz that augments the usual
power series with non-analytic factors such as
\(\exp(-kA/g)\,g^{\beta k}\).  The Schwinger–Dyson hierarchy then
becomes a triangular recursion that determines all perturbative
coefficients together with their alien derivatives, while the Stokes constants encode the genuine non-perturbative information.  In this
way the formal part \(\mathcal{M}_-\) controls large-order growth, the
analytic part \(\mathcal{M}_+\) supplies sectorial Borel–Laplace sums,
and the Stokes gluing reconstructs the full solution without truncation.

A complete analytic theory of the MS condition at infinite rank remains to be established~\cite{ecalle1981fonctions,deligne2006equations,mochizuki2011wild,kedlaya2014good}.  Outstanding tasks include controlling the infinite constellation of anti-Stokes rays generated by nested subdivergences, selecting a functional-analytic topology that makes sectorial Borel transforms converge for the full hierarchy, extending multisummability and Stokes filtration techniques to infinite rank,
and generalising wild harmonic bundle results to provide growth bounds for analytic gauges.  Meeting these challenges will turn the present geometric framework into a fully non-perturbative description of quantum field dynamics.

\newpage
\appendix

\section{Dimensional regularization and Microlocal Analysis}
\label{sec:dim_reg}

Dimensional regularization replaces the space–time dimension
\(4\) by \(d=4-\epsilon\) and multiplies the action by
the factor \(\mu^{\epsilon}\).  
Although this prescription controls ultraviolet divergences,
it breaks strict locality: the differential operators that enter the
Schwinger–Dyson (SD) equations acquire non–local kernels.
To treat the resulting equations rigorously we work in the calculus of
pseudo\-differential operators and use microlocal analysis to track
their singular structure.

\medskip
\noindent
\textbf{Regulated action and SD operator.}
For an external source \(J(x)\) the regulated action reads
\[
   S(\epsilon,\mu)
   =\mu^{\epsilon}\!
     \int d^{\,4-\epsilon}x\,
       \bigl[\mathcal{L}(\phi,g)+J(x)\phi(x)\bigr] ,
\]
with \(g\) the running coupling.
The corresponding SD operator is
\begin{equation}\label{eq:SDE_PDO}
   \mathcal{D}_{x}(\epsilon,\mu)
   =\left.\frac{\delta S(\epsilon,\mu)}{\delta\phi(x)}
     \right|_{\phi\mapsto\delta/\delta J}\!,
\end{equation}
and the SD hierarchy is
\(
   \mathcal{D}_{x}(\epsilon,\mu)\,Z[J]=0
\)
for all \(x\).
Because the variation in \eqref{eq:SDE_PDO} produces
momentum factors \(|p|^{-\epsilon}\) inside loops, each
\(\mathcal{D}_{x}(\epsilon,\mu)\) is a pseudo\-differential operator, not
a local differential one.

\medskip
\noindent
\textbf{Symbols after regularization.}
Let \(P\) denote a typical differential operator with
symbol \(\sigma_{P}(p)\).
Dimensional regularization inserts the factor
\(\bigl(\mu/|p|\bigr)^{\epsilon}\), so that, for instance,
\[
   \sigma_{\Box}(p)=|p|^{2}
   \quad\longrightarrow\quad
   \sigma_{\Box,\epsilon}(p)
   =\mu^{\epsilon}|p|^{2-\epsilon}.
\]
More generally,
\[
   P_{\epsilon}f(x)
   =\int e^{ip\cdot(x-y)}\,
        \mu^{\epsilon}\sigma_{P}(p)|p|^{-\epsilon}\,f(y)\,d^{4-\epsilon}p .
\]
The modified symbol belongs to the same H\"ormander class
\(S^{m-\epsilon}_{1,0}\) as the original one of order \(m\),
hence \(P_{\epsilon}\) remains a bona fide pseudo\-differential operator.

\medskip
\noindent
\textbf{Microlocal properties.}
The symbol
\(\mu^{\epsilon}|p|^{2-\epsilon}\) is smooth in \(p\) for
\(\epsilon\) near \(0\), homogeneous of degree \(2-\epsilon\),
analytic in \(\epsilon\) and tempered both at
\(p=0\) and at large \(|p|\).
Consequently
\begin{itemize}
\item the wavefront set of \(P_{\epsilon}f\) is contained in that of
      \(f\), so regularization does not create new space–time
      singularities,
\item the class of microlocal D-modules is preserved: every
      \(\mathcal{D}_{x}(\epsilon,\mu)\) acts continuously on the
      same spaces of distributions as before regularization.
\end{itemize}

\medskip
\noindent
\textbf{Consistency for the  Schwinger–Dyson hierarchy.}
The regularizing factor
\(\bigl(\mu/|p|\bigr)^{\epsilon}\) multiplies every loop integration
kernel, so \emph{every} \(n\)-point equation in the hierarchy is
governed by operators whose symbols differ from their four–dimensional
counterparts by exactly the same analytic factor
\(|p|^{-\epsilon}\mu^{\epsilon}\).
The pseudo\-differential calculus is closed under such scalar
multiplications; compositions, adjoints and commutators stay within the
same H\"ormander class.
Consequently the standard microlocal filtration that splits a symbol
into its local part (polynomial in \(p\)) and its non–local analytic
tail is preserved at each order.
The full tower of Schwinger–Dyson equations therefore forms a single,
well–defined microlocal D-module: a sheaf whose sections are
distributions with controlled wavefront sets and whose structure maps
are given by the uniformly regularized pseudo\-differential operators
\(\mathcal{D}_{x}(\epsilon,\mu)\).

\section{Perturbatively renormalizable QFTs satisfy the finite rank condition}
\label{sec:finite_rank_perturbative}

Consider a \(D\)-dimensional perturbatively renormalizable quantum field theory whose bare
Lagrangian contains only operators of engineering dimension not
exceeding \(D\):
\[
   \mathcal{L}_{\text{bare}}
   =\sum_{i=1}^{N}g_{0\,i}\,\mathcal{O}_{i}(x),
   \qquad
   \dim\mathcal{O}_{i}\le D .
\]
Locality and power counting imply that ultraviolet divergences can
renormalize no operators beyond this finite list, hence the index
\(i\) takes just \(N\) values.  Minimal subtraction introduces the same
operators as counterterms through the relations
\(g_{0\,i}=Z_{ij}\,g_{j}\mu^{\epsilon}\) with a mixing matrix
\(Z_{ij}=\delta_{ij}+O(\hbar)\) of finite size.

Collecting the bare couplings in the column
\(g_{0}=(g_{0\,1},\ldots,g_{0\,N})^{\top}\) the renormalization group
acts according to
\[
   \mu\frac{\mathrm{d}}{\mathrm{d}\mu}g_{i}
   =\beta_{i}(g)
   =\sum_{j=1}^{N}A_{ij}(g)\,g_{j},
   \qquad
   A_{ij}(g)=\frac{\partial\beta_{i}}{\partial g_{j}} ,
\]
so the scale connection involves the finite matrix \(A(g)\).
The anomalous dimensions form another finite matrix
\(\gamma_{ij}=-(\mu/2)\,\partial_\mu\ln Z_{ij}\).

Introduce the vector of renormalized Green functions
\[
   X(x_{1},\ldots,x_{n})
   =\bigl(
        \langle\mathcal{O}_{1}(x_{1})\cdots\rangle,\,
        \ldots,\,
        \langle\mathcal{O}_{N}(x_{1})\cdots\rangle
     \bigr)^{\top}.
\]
Schwinger–Dyson identities, the Callan–Symanzik equation and the renormalization group give 
\[
   \bigl(\partial_\epsilon+A(\epsilon,\mu)\bigr)X=0,
   \qquad
   \bigl(\mu\partial_\mu+\gamma(\mu)\bigr)X=0 ,
\]
with \(A(\epsilon,\mu)\) and \(\gamma(\mu)\) both acting on the
finite-dimensional space \(\mathbb{C}^{N}\cong
\operatorname{span}\{\mathcal{O}_{1},\ldots,\mathcal{O}_{N}\}\).
The associated D-module
\(\mathcal{M}=\operatorname{span}\{X\}\) therefore has rank
\(N<\infty\).

Because all divergent operators and couplings fit inside this finite
vector space every perturbatively renormalizable local quantum field
theory satisfies the finite rank condition.  The classical
Levelt–Turrittin and Malgrange–Sibuya theorems then apply without
modification, providing the analytic gauge together with the standard
perturbative renormalization framework.


\bibliographystyle{plain}
\bibliography{ref}

\end{document}